\documentclass{article}
\usepackage{arxiv}
\usepackage{hyperref}
\usepackage{latexsym}
\usepackage{url}
\usepackage[utf8]{inputenc}
\usepackage{booktabs}
\usepackage{amsmath}
\usepackage{amsfonts}
\usepackage{amssymb}
\usepackage{longtable}
\usepackage[inline]{enumitem}
\usepackage{enumitem}
\usepackage{mathbbol}
\usepackage[T1]{fontenc}
\usepackage{mathtools}
\usepackage{multirow}
\usepackage{natbib}
\usepackage{rotating}
\usepackage{subcaption}
\usepackage{graphicx}
\usepackage{etaremune}
\usepackage{authblk}
\usepackage{tikz}

\title{Overview of the TREC 2025 Million Large Language Models track}

\author[1]{Evangelos Kanoulas}
\author[1]{Panagiotis Eustratiadis}
\author[2]{Jamie Callan}
\author[3]{Mark Sanderson}
\author[1]{Yongkang Li}
\author[1]{Jingfen Qiao}
\author[1]{Gabrielle Poerwawinata}
\author[1]{Vaishali Pal}

\affil[1]{University of Amsterdam, Netherlands\\\texttt{\small \{e.kanoulas,p.efstratiadis,y.li7,j.qiao,g.poerwawinata,v.pal\}@uva.nl}}
\affil[2]{Carnegie Mellon University, USA\\\texttt{\small callan@cs.cmu.edu}}
\affil[4]{RMIT, Australia\\\texttt{\small mark.sanderson@rmit.edu.au}}

\begin{document}

\maketitle

\begin{abstract}
Agentic AI envisions ecosystems of intelligent agents collaboratively solving complex tasks with minimal human intervention. In such ecosystems, each agent possesses specialized expertise, making effective expert selection central to overall system performance. While most current approaches assume a small number of well-documented models, real-world expertise is far more diverse and cannot be adequately captured through static metadata or hand-written descriptions.

We anticipate a future with millions of specialized language models (LLMs), each excelling in different domains or problem types. Rather than relying on predefined capability statements, we propose a retrieval-based paradigm in which an assistant agent infers expertise dynamically by examining models’ observable behavior. Upon receiving a user query, the assistant ranks candidate LLMs based on demonstrated competence, enabling efficient and adaptive expert selection.

The TREC Million LLM Track operationalizes this paradigm by shifting the retrieval target from documents to expert LLMs. Participants are given a discovery set consisting of queries, answers, and log-probabilities from more than one thousand LLMs, and are challenged to infer meaningful expertise representations for each model. Given an unseen test query, systems must then rank the LLMs according to their expected performance, providing the first large-scale benchmark for expertise retrieval in agentic AI.
\end{abstract}

\begin{center}
    \textbf{Track website:} \url{https://trec-mllm.github.io/}
\end{center}

\section{Introduction}

The rapid advancement of generative artificial intelligence has significantly reshaped the field of information retrieval (IR)~\cite{zhu2023large,allan-2024-future,white2025information}. Traditionally, a user-provided query prompts an IR system to align the query's intent with a pre-existing corpus of textual resources, returning a ranked list of documents from which the user then extracts relevant information. In this paper, we envision a future beyond traditional information access, in which information is served to users through (large) language models (LLMs)~\cite{openai2022chatgpt, wei2022emergentabilitieslargelanguage}. As generative AI models increasingly act as primary sources of information, this shift necessitates a new paradigm. This paradigm must address the challenge of selecting the most suitable LLMs to query and synthesizing knowledge across multiple expert models.

This future is unfolding at a fast pace~\cite{caramancion2024large}\footnote{\url{https://kafkai.com/en/blog/ai-impact-on-search-engines/}}, driven by the proliferation of domain-specialized LLMs tailored to increasingly narrow fields of expertise. We are moving toward a world where many  such expert LLMs will exist, each trained on distinct knowledge domains or accessing distinct databases (RAG ), some overlapping, others highly specialized. This shift is already underway, with companies like BloombergGPT~\cite{wu2023bloomberggpt}, Thomson Reuters\footnote{\url{https://legalsolutions.thomsonreuters.co.uk/en/c/practical-law/now-with-generative-ai.html}}, Elsevier\footnote{\url{https://www.elsevier.com/products/scopus/scopus-ai}} developing LLMs on proprietary material. 
As this paradigm evolves, a user's information needs will increasingly be met not by a single system, but by querying and synthesizing responses from multiple expert LLMs, requiring careful orchestration to extract the most accurate and relevant insights.


In this evolving digital ecosystem, an agent will be necessary to orchestrate queries and synthesize responses from domain-expert LLMs. However, querying all available LLMs is prohibitively expensive and inefficient. Thus, a new type of search engine is required that, given a user's question, intelligently retrieves only the most relevant expert LLMs. This agent must dynamically determine which expert models to consult, balancing the need for accuracy with cost efficiency. Upon receiving a user's query, it should strategically decide which LLMs to engage with, ensuring the response is both comprehensive and cost-effective while minimizing redundant computations.

In the TREC Million LLM Track, we take what we believe is the first concrete step toward building such a framework. Rather than retrieving documents or passages, the track shifts the retrieval target to expert models themselves. The goal is to evaluate systems on their ability to identify which LLMs are most capable of answering a new query, based not on stated capabilities or fixed metadata, but on observable behavior. This reframes the classic retrieval problem: instead of mapping queries to documents, we now map queries to models.

To support this new retrieval paradigm, the track provides a large and diverse pool of more than one thousand LLMs, spanning different sizes, training regimes, and domains of apparent expertise. Crucially, participants are not given high-level descriptions of these models. Instead, each LLM is represented by a discovery set consisting of hundreds of queries paired with the model’s generated answers and token-level log-probabilities. These data serve as behavioral evidence from which participants must infer each model’s strengths, weaknesses, and characteristic expertise patterns.

Given an unseen test query, a participating system must then rank all LLMs according to their expected competence.
\section{Task Description}

The TREC 2025 Million LLMs (MLLM) Track addresses the problem of \emph{expert model retrieval}: given a user query, the goal is to predict which language models (LMs) are most capable of generating a high-quality answer. Unlike traditional information retrieval tasks that focus on ranking documents or passages, participants must rank a fixed pool of LMs according to their expected performance on each query.

Participants are provided with three key resources:
\begin{itemize}
    \item \textbf{A set of LM identifiers} corresponding to more than one thousand LMs.
    \item \textbf{A discovery dataset} containing previously collected responses from all LMs on a large set of queries. Each response includes both the generated text and token-level log-probabilities, enabling participants to infer behavioral and stylistic characteristics of each model.
    \item \textbf{A development set} consisting of queries paired with ground-truth expertise labels indicating which LMs produced the best answers for those specific queries.
\end{itemize}

Using only these resources---and without issuing any new queries to the LMs during evaluation---participants must build systems that learn representations of LM expertise and use them to predict model competence on unseen test queries. For each query in the test set, a run must output a ranked list of all LM identifiers, ordered from most to least likely to produce a high-quality answer.

Submissions are evaluated against hidden queries and corresponding ground-truth expertise labels derived from human and LM judgments. We report performance using standard ranking metrics, including nDCG@10 and MRR, which capture both the accuracy and the ordering quality of predicted expert rankings.

This formulation encourages systems that can generalize from observable model behavior to latent expertise, providing the first large-scale benchmark for expertise retrieval in agentic AI.

\section{Simulating Thousands of Expert LLMs}
\label{sec:simulation}

\begin{figure}[t]
    \centering
    \includegraphics[width=0.8\linewidth]{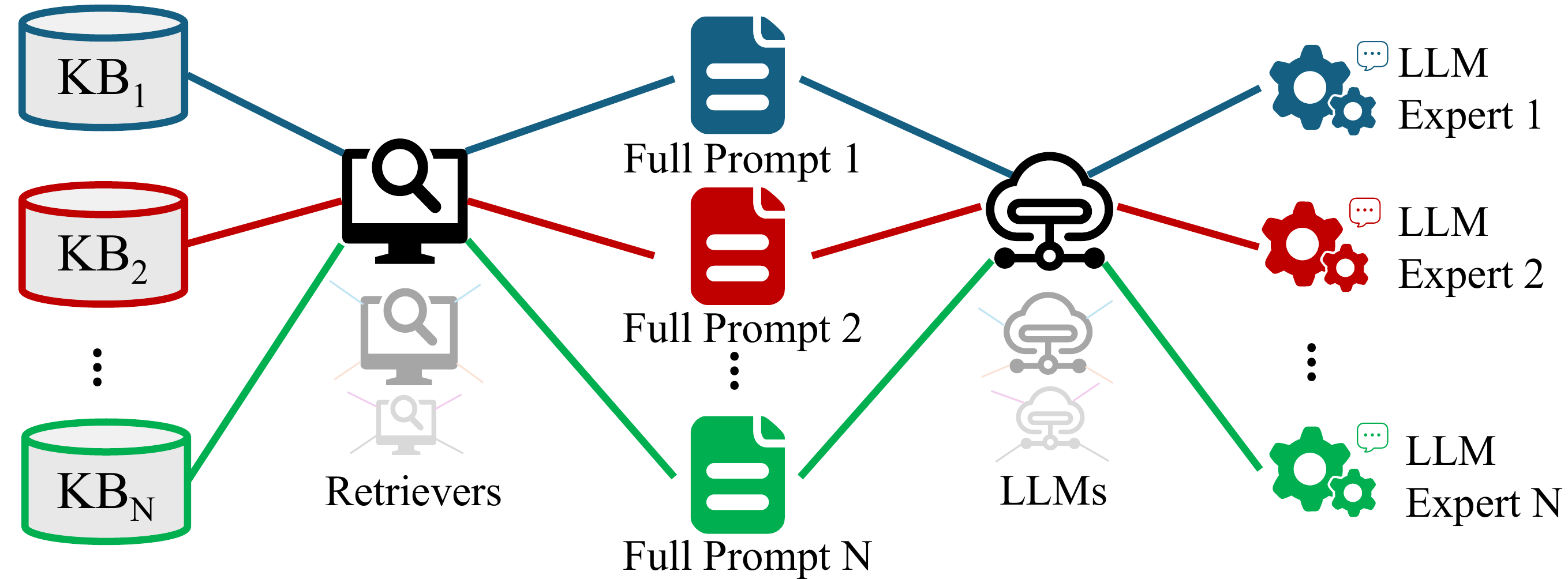}
    \caption{Using RAG-based LLMs to simulate expert models. Each expert consists of a shared LLM and retriever, but is restricted to a distinct document collection---typically a structured KB---that defines its domain expertise.}
    \label{fig:simulation}
\end{figure}

As described earlier, the Million LLM Track does not provide direct access to proprietary LLMs, but instead provides their responses. This raises a natural question: \emph{how can we obtain responses from thousands of domain-specialized LLMs?} In practice, such a massive collection of expert LLMs—spanning sports, finance, medicine, music, cinema, and more—does not yet exist. To approximate this setting, we simulate expert models using a retrieval-augmented generation (RAG) paradigm, where each simulated expert consists of:
\begin{itemize}
    \item a \textbf{shared underlying LLM} responsible for generating answers,
    \item a \textbf{dedicated retriever} for retrieving relevant documents, and
    \item a \textbf{unique document collection} defining the expert's domain.
\end{itemize}

Under this setup, an LLM paired with a cinema-related document collection behaves as a \emph{cinema expert}, while another equipped with financial reports behaves as a \emph{finance expert}. By restricting retrieval to domain-specific documents, each simulated expert is constrained to operate within a well-defined knowledge scope, approximating the behavior of real-world specialized LLMs (Figure~\ref{fig:simulation}).

\subsection*{Preventing Knowledge Leakage}

A key challenge in simulating expert LLMs is preventing leakage from the shared base model. Ideally, experts should rely only on retrieved documents, not on the model's parametric memory. In practice, however, fully disentangling generation ability from pre-training knowledge is impossible.

To mitigate leakage, we employ two complementary strategies:

\paragraph{(i) Prompt Engineering.}  
We explicitly instruct the model to base its answers solely on the retrieved documents. This sometimes results in responses such as ``No result found,'' indicating that the model adhered to instructions. However, previous work shows that instruction-following is imperfect, and LLMs often introduce additional pre-trained knowledge.

\paragraph{(ii) Query Filtering.}  
A more robust strategy is to filter out queries whose correct answers are already known to the base model. For collections with ground-truth answers, we compare the base model's response to the ground truth; queries the base model answers correctly are removed.  
In collections lacking explicit ground truth but providing relevant documents, we compare the base model's response with the retrieved evidence and discard queries for which the model's answer is entailed by the retrieved documents.  
In both cases, retained queries force the model to rely on retrieval rather than prior knowledge, producing high-quality training data for simulated experts.

\subsection*{Constructing Thousands of Expert Domains}

Ideally, each expert domain would have its own curated, domain-specific document collection. However, constructing thousands of such collections manually is impractical. Instead, we reuse existing large-scale document datasets. For this approach to be effective, the chosen dataset must:
\begin{itemize}
    \item be sufficiently large to support thousands of meaningful sub-collections,
    \item provide a substantial set of queries, either with ground-truth answers or with associated relevant documents.
\end{itemize}

Given such a dataset, we partition the entire collection into thousands of clusters, each representing a specialized domain. For each simulated expert model, retrieval is limited to documents within its assigned cluster; the retrieved documents are then used as input context for the shared LLM to generate responses. This yields thousands of domain-constrained expert models.

\subsection*{Efficient Retrieval at Scale}

A naïve implementation would require building a separate index for each cluster, which is computationally infeasible at the scale of thousands of experts. Instead, we adopt a more efficient strategy using a single global index:
\begin{enumerate}
    \item Retrieve the top-$k$ documents from the global index.
    \item Identify which clusters these documents belong to.
    \item For each cluster represented among the top-$k$, forward the top-$c$ documents to the LLM to generate the cluster-specific response.
    \item For clusters absent from the top-$k$, provide a randomly sampled document from the cluster, since the entire cluster is likely irrelevant to the query.
\end{enumerate}

This approach preserves the specialization of each expert while avoiding the overhead of maintaining thousands of separate search systems. It enables the scalable simulation of a large population of expert LLMs suitable for benchmarking expertise retrieval at TREC scale.

We further sampled 342 additional queries from the researchy questions, drawn from the \href{https://huggingface.co/datasets/corbyrosset/researchy_questions}{Researchy Questions} dataset, a question–answering benchmark built on top of ClueWeb22 with non-factoid researchy queries. For each one of the additional queries we also obtained the golden answers in the researchy questions dataset. We then used GPT-o1 as LLM-as-a-Judge and developed a 3-level qrels comparing the LLMs' answers to the golden answer. We released this dataset as a development set.

A set of testing queries were developed by NIST assessors, who had access to the collection but also access to answers from gpt4o, gpt5-nano, and phi4, so that they can generate questions for which these models did not provide a good answer but such an answer was in the collection. NIST provided us with 50 such questions. For the evaluation the goal was to pool answers from the top-ranked by participant LLMs. NIST assessors judge these answers to generate relevance labels for LLM expertise. Unfortunately, the plan was able to be implemented. Instead we collected ground truth answers for 37 out of the 50 questions. Using GPT5-as-a-judge we compared the ground truth answer with the answers from all LLMs and generated a 3-grained expertise labels.

\section{Implementation Details}

\paragraph{Document Collection}
To support the evaluation of multi-LLM retrieval systems, we curated a large-scale, semantically enriched document collection derived from the ClueWeb22 corpus. Our final corpus is constructed by combining and processing two ClueWeb22-based sources.

First, we start from the MS MARCO Web Search task, which released two versions of its corpus: one based on the full ClueWeb22-L collection and a more curated 200-million-document subset. We use the latter and retain only English-language documents from the ClueWeb22-A portion, resulting in 69,947,610 documents.

Second, we collect all ClueWeb22 documents referenced in Researchy Questions dataset, yielding an additional 332,874 documents.

Rather than clustering documents directly based on their raw text, we cluster them using semantic category tags provided by Bing. These tags are available only for \href{http://boston.lti.cs.cmu.edu/callan/Data/cw22-topics/}{ClueWeb22-A} and, more specifically, for the English shards en01–en09, covering 812,572,651 valid topic annotations. We intersect this topic file with our two source corpora. For the MS MARCO Web Search–derived corpus, this matching step yields 26,426,735 documents for which at least one Bing tag is available. Applying the same procedure to the researchy questions corpus yields 146,644 tagged documents.

We then take the union of these two tagged corpora, remove duplicate documents, and filter out pages with empty or missing content. After these cleaning steps, we obtain a final corpus of 26,566,121 documents. This semantically tagged ClueWeb22-derived collection serves as the basis for our subsequent clustering and multi-LLM retrieval experiments.

\paragraph{Clustering} Clustering has two stages:
\begin{enumerate}
\item Initial label assignment: We use K-means and the Bing topic tags as sparse vectors and do the initial clustering, creating 1500 clusters. At this stage the 1500 clusters are really imbalanced.
\item Cluster balancing:
    \begin{enumerate}
    \item We split really large clusters (cluster size > max size) into several clusters of size max size. We use max size = 100K documents.
    \item We merge really small clusters (cluster size < min size) into clusters of roughly size min size. We use min size = 1000 documents.
    \end{enumerate}
\item We verify that every original cluster whose size is in [min size, max size] remains intact (i.e., all its members map to a single new label, and that new label's cluster has exactly the same size.
\end{enumerate}

\begin{figure}[t]
    \centering
    \includegraphics[width=1\linewidth]{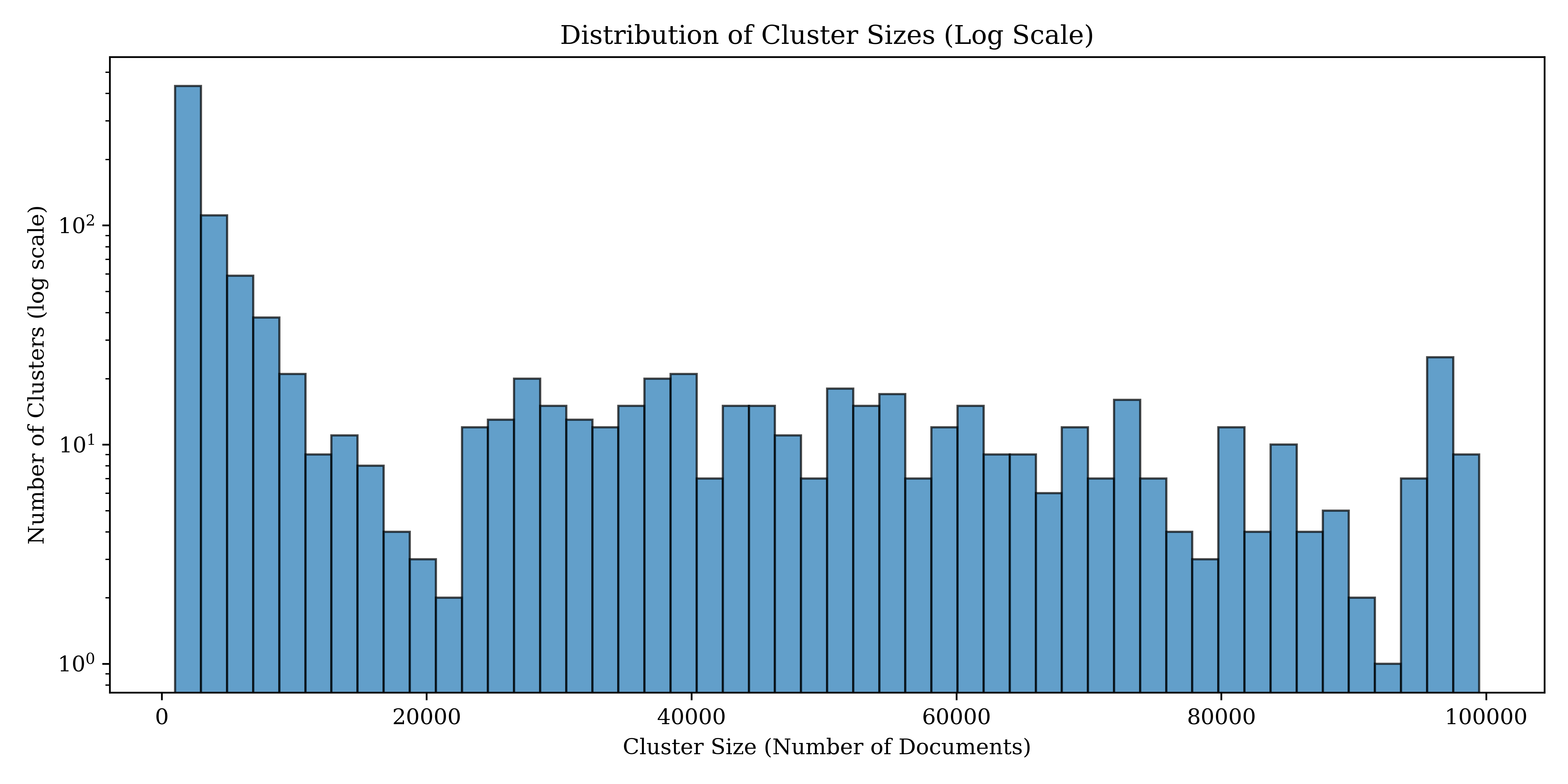}
    \caption{Distribution of cluster sizes across all document clusters. The y-axis is shown on a logarithmic scale.}
    \label{fig:cluster_size_distribution}
\end{figure}

\begin{figure}[t]
    \centering
    \includegraphics[width=1\linewidth]{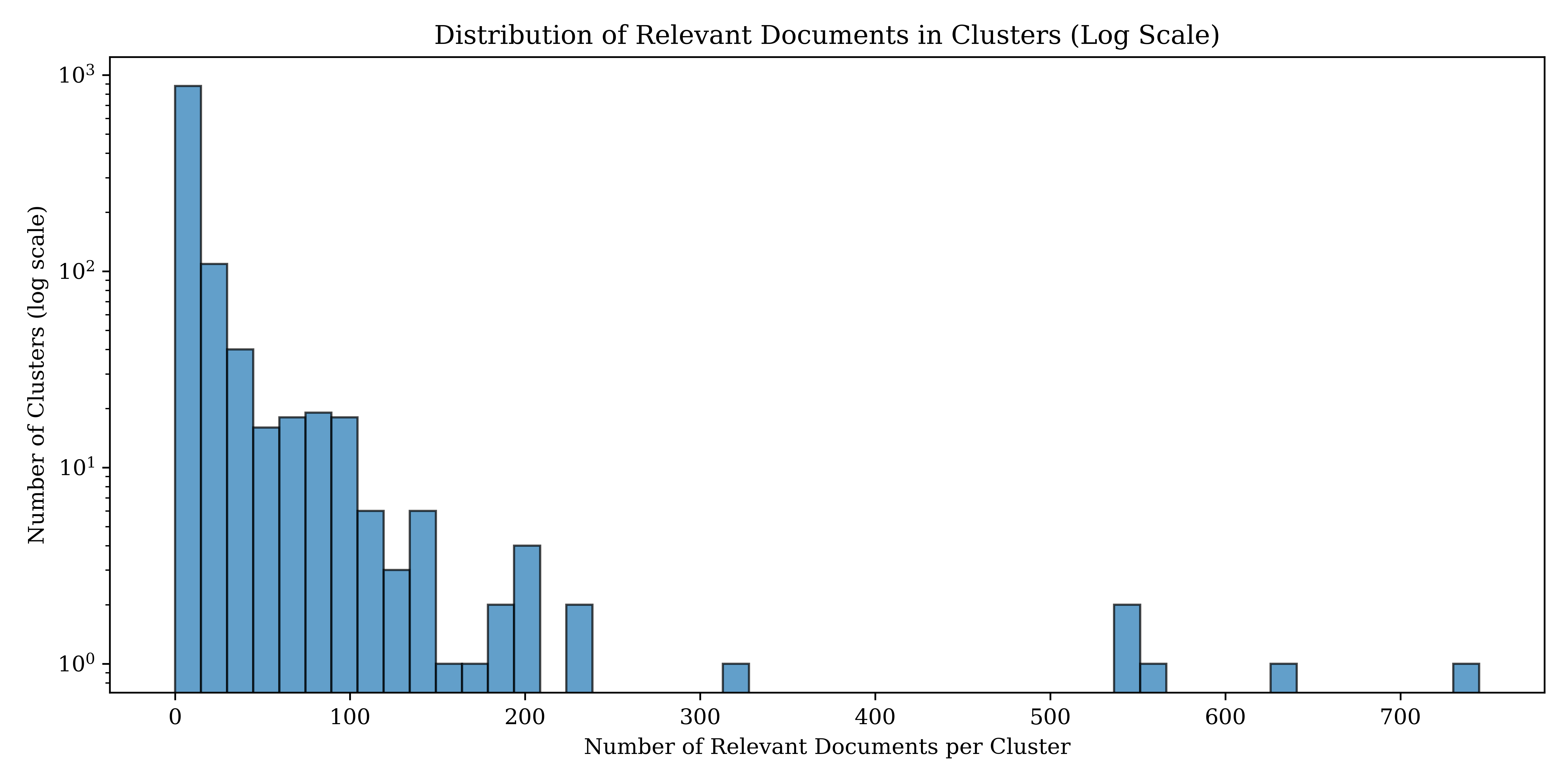}
    \caption{Distribution of relevant documents across clusters. The y-axis is shown on a logarithmic scale.}
    \label{fig:cluster_ground_truth_distribution}
\end{figure}

\paragraph{Retrieval} Ideally, we would simulate expert LLMs by building 1{,}131 Retrieval-Augmented Generation (RAG) systems, each consisting of a dedicated index over its cluster’s documents, a retriever, and a shared LLM backend. While this setup provides the desired specialization, it is computationally inefficient, as it requires maintaining 1{,}131 separate indexes and querying all systems for every input, even though most clusters are irrelevant to any given query.

To improve scalability, we implemented a more efficient alternative: a single global index is built over the entire 80.5 million-document corpus. For each incoming query, the system:
\begin{enumerate}
    \item retrieves the top 2{,}000 documents globally using BM25;
    \item for each of the 1{,}131 clusters, selects up to the top 3 documents from that cluster among the global top 2{,}000;
    \item passes these documents to the shared LLM to simulate a cluster-specific response;
    \item returns a default template response for clusters that have no documents in the global top 2{,}000.
\end{enumerate}
This optimization retains the domain specialization of the simulated experts while enabling efficient, large-scale evaluation.

\paragraph{Query Collection}
To support robust evaluation of expert LLM selection, we curated a balanced set of user questions drawn from two distinct and complementary subsets of the ClueWeb22 ecosystem. These subsets differ in both query style and information complexity, enabling us to assess retrieval effectiveness across a wide range of user intents.

The first 9{,}989 queries were selected from the MS MARCO Web Search dataset, which contains over 10 million real Bing queries paired with positively clicked documents. These queries are typically short, under-specified, and keyword-style, reflecting common search engine behavior. To ensure that responses to these queries genuinely require external knowledge retrieval, we implemented a filtering mechanism. For each candidate query, we compared the LLM’s answer with and without access to the clicked document. Specifically, we prompted the LLM with (i) only the query and (ii) the query plus the associated clicked document. Using GPT-4o as an LLM-as-a-Judge, we retained only those queries for which the model's answer changed meaningfully when retrieval was introduced. This process ensured that the final query set could not be trivially answered from the LLM’s internal memory and required grounding in the document collection.

The second 4{,}961 queries were sampled from a pool of 96{,}450 researchy natural language questions, also sourced from Bing query logs. We retained only queries whose user-clicked documents appear in the 1{,}113 clusters defined above. These queries were selected based on signals such as long user dwell time and richer engagement, indicating more exploratory or in-depth information needs. Unlike the terse, keyword-based queries in the first set, these are fully formed natural language questions, often complex or multi-faceted. Due to their inherently open-ended structure, we assume that general-purpose LLMs would struggle to answer them accurately without retrieval, and thus we sampled them randomly without additional filtering. Together, these two query sets offer a challenging and diverse testbed for benchmarking expert LLM selection across both entity-centric and knowledge-intensive information needs.

\paragraph{Development Set}
We further sampled 342 additional queries from the Researchy Questions dataset. For each of these queries, we also obtained the gold answers provided by the dataset. We then used GPT-o1 as an LLM-as-a-Judge to compare each LLM’s answer to the gold answer and constructed three-level relevance judgments (qrels) reflecting different degrees of expertise. We release this dataset as a development set.

\paragraph{Test Set}
The test queries were developed by NIST assessors, who had access to the document collection as well as answers from GPT-4o, GPT-5-nano, and Phi-4. Their task was to generate questions for which these models did not provide a good answer, even though a good answer existed in the collection. NIST provided 50 such questions.

Our original evaluation plan was to pool answers from the top-ranked LLMs (as predicted by participants’ systems) and have NIST assessors label these answers to derive expertise judgments. However, this plan was not fully implemented. Instead, NIST provided ground truth answers for 37 out of the 50 questions and used GPT-5-as-a-Judge to compare these answers with the responses from all LLMs, generating three-graded expertise labels for evaluation. Out of these 37 questions only 16 appeared to have evoked at least one relevant answer from the LLMs.

The prompt is as follows:
\begin{verbatim}
user_context = (
        "You are a helpful assistant. You will receive a question, a LLM answer, 
        and a ground truth answer. "
        "Your task is to judge whether the LLM answer can answer the question 
        properly compare to the ground truth answer. "
        "If the LLM answer can answer the question properly, you should return 2, 
        as this LLM is an expert in this question. "
        "If the LLM know part of the question, you should return 1, as this LLM is 
        not an expert but can answer part of the question. "
        "If the LLM know nothing about the question, you should return 0, as this 
        LLM is not an expert in this question. "
        "In general, I divide the scores into three levels: 0, 1, and 2, which 
        respectively represent LLMs with zero knowledge, have some knowledge, and 
        are experts."
        "Please directly output the corresponding scores, do not add any other 
        information. \n"
         "Question: {query}\n"
         "LLM Answer: {llm_answer}\n"
         "Ground Truth Answer: {ground_truth_answer}\n"
    )

messages = [
{"role": "system", "content": "You are a knowledgeable assistant, your task is to 
judge whether an LLM can answer the question properly compare to the ground truth 
answer I give you. "},
{"role": "user", "content":  user_context},
]
\end{verbatim}
\section{Summary of Participating Systems}

A total of 19 runs were submitted by six participating groups. Despite the novelty of the task—ranking \emph{language models} rather than documents—participants explored a wide range of approaches, from classical IR baselines and feature-driven machine learning to modern transformer-based learning-to-rank pipelines and LLM-assisted meta-evaluation. Below, we summarize the main methodological directions pursued by each group.

\paragraph{GenAIus} GenAIus formulated the task as an expert retrieval problem and adapted classical expert-finding methods to rank LLMs. They explored both profile-based and document-based approaches. In the profile-based variant, each LLM was represented by a single aggregated profile built from all of its query--response pairs, which was then ranked directly using BM25. In the document-based variants, either individual query--response pairs or query-level documents were indexed, and model rankings were produced by aggregating evidence from retrieved documents using expert-search voting models such as \textsc{Vote} and \textsc{SumScore}. They also introduced a two-stage filtering pipeline that removed ``No result found'' outputs and low-confidence answers, and experimented with RM3-based query expansion and \texttt{naver-splade-v3} reranking. Their best-performing submissions were response-based systems, showing that fine-grained response-level evidence was more effective than profile-level or question-level representations for LLM expertise estimation.

\paragraph{SRCB} SRCB approached the task as a supervised ranking problem under weak or automatically constructed labels. To address the absence of ground-truth labels in the discovery set, they generated training supervision using several LLM-based annotation strategies, including a Quicksort-style pairwise ranking procedure, a reranker-based scoring method, and embedding-similarity approaches, as well as combinations of these signals. On top of these labels, they explored multiple learning formulations---ranking, classification, and regression---using a pretrained encoder plus lightweight task-specific decoder architecture, with the encoder frozen for efficiency at Million-LLM scale. Their experiments compared different encoders, including \texttt{DeBERTa-v3-base}, \texttt{Qwen3-Embedding-8B}, and \texttt{Qwen3-32B}, and showed strong performance from Qwen-based embedding models. Overall, SRCB’s submissions highlight the effectiveness of LLM-based pseudo-label construction and efficient supervised neural ranking for LLM selection.

\paragraph{UDInfo} UDInfo explored a diverse set of embedding-based routing methods for LLM selection, ranging from unsupervised profile construction to weakly supervised neural routing. Their strongest approach was a hierarchical two-stage profile model that first represented each LLM through clusters of historical answer embeddings and used these cluster centroids for broad scoring, then refined the ranking with per-model nearest-neighbor search over answer embeddings. They also investigated a hybrid query-space method that used weak supervision from an external LLM judge to estimate model expertise through neighborhood-based and cluster-based query profiles, as well as simpler centroid-profile baselines and multi-class neural routers implemented as shallow and deep MLP classifiers over query embeddings. Across these variants, their results suggested that retrieval-style profile methods in embedding space were more effective than purely parametric classifiers, highlighting the value of combining global expertise profiles with local query-specific evidence.

\paragraph{IndeLab} IndeLab provided the broadest variety of runs, including both sophisticated neural systems and interpretable non-neural baselines. Their neural submissions incorporate task-specific sub-prompt classifiers, feature-based score computation modules, and aggregation layers that combine multiple signals into final rankings. The group also submitted a non-neural ensemble method using rank-fusion techniques, offering a classical IR reference point for the task. Collectively, IndeLab’s submissions illustrate a spectrum of lightweight and modular approaches to the LLM-ranking problem.

\paragraph{paulphys} The paulphys group contributed systems based on listwise learning-to-rank, directly optimizing ranking losses over discovery-set features and model metadata. Their submissions implement focused neural LTR architectures tailored specifically to the Million-LLM setting.

\paragraph{UvA IRLab} UvA IRLab approached LLM selection as a supervised learning-to-rank problem using next-token log probabilities from the discovery set as zero-cost pseudo-relevance signals. For each query--LLM pair, they derived soft labels from the mean log-probability of the model’s generated response, treating model confidence as a proxy for answer quality. Their main system trained a LightGBM LambdaRank model over feature vectors combining query embeddings, LLM identifiers, and global per-model statistics computed from top-percentile log-probability scores. They also evaluated unsupervised tag-based and response-based baselines, and conducted ablations showing that global LLM-level statistics were stronger predictors than query-specific semantic features in the current setting. In post-submission analysis, they further extended the model with topic-aware statistics obtained by clustering queries, allowing the ranker to better capture domain-specific strengths of different LLMs.

\subsection*{Use of Metadata, External Models, and Development Data}

We observe heterogeneous usage of the additional data resources provided by the track. Only three runs made use of the \texttt{mllm-metadata} package—which includes per-model behavioral statistics such as log-probability summaries—while sixteen runs did not incorporate this information. External models were used more frequently: four runs employed an external LLM (e.g., for annotation or feature extraction), whereas fifteen runs relied exclusively on the provided resources.

All nineteen submissions used the \texttt{mllm-devset}, underscoring its centrality for understanding model behavior on held-out queries. However, only six runs used this devset for training purposes (\texttt{mllm-devtrain = yes}), typically in supervised learning-to-rank settings or classifier-based pipelines. The remaining thirteen runs used the devset exclusively for validation, feature extraction, system calibration, or hyperparameter tuning. These patterns reflect the diversity of methodological philosophies across participating teams, ranging from feature-driven and heuristic approaches to fully supervised neural pipelines with LLM-generated training signals.

\section{Results}

Table~\ref{tab:ndcg10-extended} presents the nDCG@10 averaged over 16 scores for all submitted runs, alongside indicators for whether a system used (i) external models and (ii) the development set for training.

\paragraph{Top-performing systems.}
The strongest runs achieved nDCG@10 scores between 0.15 and 0.17. The best-performing system, \texttt{submission\_4} (0.17447), was followed closely by \texttt{infolab\_UD\_run2} (0.16913) and \texttt{r-ef-rm3-1000-ss} (0.16801). These systems were predominantly based on unsupervised techniques. Their strong performance suggests that behavioral signals contained within the model-generated responses---such as answer structure, lexical cues, or implicit quality indicators---are highly predictive of future LLM expertise.

\paragraph{Effect of devset training.}
Only six submissions made use of the development set for supervised training. Among these, the \texttt{ensemble\_v1}, \texttt{ensemble\_v2}, and \texttt{ensemble\_v3} runs were the most competitive, achieving scores between 0.147 and 0.152. These results indicate that supervised training can offer benefits. Other devset-trained runs, such as \texttt{mllm-indelab-09-17} and \texttt{lambdamart\_profiles}, performed worse. This likely reflects the limited size of the devset (342 queries) and the difference in distribution between the development and test queries, both of which constrain generalization.

\begin{table}[t]
\centering
\caption{NDCG@10 scores with external model usage and devset training information.}
\label{tab:ndcg10-extended}
\begin{tabular}{l r c c}
\hline
\textbf{System} & \textbf{NDCG@10} & \textbf{External Models} & \textbf{Devset Training} \\
\hline
submission\_4 & 0.17447 & no & no \\
infolab\_UD\_run2 & 0.16913 & no & no \\
r-ef-rm3-1000-ss & 0.16801 & no & no \\
r-f-rm3-1000-rr-ss & 0.16081 & no & no \\
r-f-rm3-1000-ss & 0.15446 & no & no \\
infolab\_UD\_run4 & 0.15341 & no & no \\
ensemble\_v1 & 0.15215 & no & yes \\
infolab\_UD\_run1 & 0.15168 & no & no \\
ensemble\_v3 & 0.14851 & no & yes \\
ensemble\_v2 & 0.14772 & no & yes \\
submission\_5 & 0.13388 & no & no \\
non-neural & 0.13091 & no & no \\
lightgbm\_job266431 & 0.12326 & no & yes \\
q-f-rm3-100-ss & 0.12030 & no & no \\
llm-f-100 & 0.10596 & yes & no \\
infolab\_UD\_run5 & 0.10206 & no & no \\
infolab\_UD\_run3 & 0.10178 & no & no \\
mllm-indelab-09-17 & 0.08140 & no & yes \\
lambdamart\_profiles & 0.05490 & yes & yes \\
\hline
\end{tabular}
\end{table}
\section{Conclusion}

The emergence of multi-LLM retrieval represents a fundamental shift in information access, moving from document ranking toward the selection and orchestration of expert models. This transition challenges long-established IR assumptions, demanding new methodologies for expertise identification, evaluation, and cost-aware decision making. As AI ecosystems grow toward millions of specialized LLMs—each with distinct domain expertise, behavior, and reliability—the need for scalable mechanisms to retrieve, compare, and select among these models becomes increasingly critical.

This was the first run of the track, we faced challenges in establishing the datasets for our task. Results were based on a small number of test set queries (16) and no human relevance judgments. The TREC Million LLMs Track provides a first step toward this future by reframing retrieval around \emph{expert model ranking}. Through a large-scale simulation of more than one thousand domain-specialized LLMs—each constrained to its own document cluster—we created a controlled environment for studying how systems can infer expertise from behavioral evidence. The discovery set, development set, and test queries collectively enabled participants to explore a diverse range of ranking strategies, from classical IR heuristics and feature-driven machine learning to supervised neural architectures and LLM-assisted meta-evaluation.



\section*{Acknowledgments}

This work was partially supported by the China Scholarship Council (Grant No.\ 202308440220), the LESSEN project (NWA.1389.20.183) of the research program NWA ORC 2020/21 funded by the Dutch Research Council (NWO), and the PACINO project (Grant No.\ 215742) financed by the Swiss National Science Foundation (SNSF).



\bibliographystyle{ACM-Reference-Format}
\bibliography{references}


\end{document}